\documentclass[prl,twocolumn,superscriptaddress]{revtex4}
\usepackage{epsfig}
\usepackage{times}
\usepackage{amsmath}
\usepackage{amsfonts}
\usepackage{amssymb}
\usepackage[usenames]{color}
\usepackage{graphicx}
\newcommand{\NTT}{NTT Basic Research Laboratories, NTT Corporation, 3-1 Morinosato-Wakamiya, Atsugi, Kanagawa, 243-0198, Japan.}

\newcommand{\NICT}{National Institute of Information and Communications
Technology, 4-2-1,
Nukuikitamachi, Koganei-city, Tokyo 184-8795 Japan}
\newcommand{\TSUKUBA}{Graduate School of Pure and Applied Sciences, University of Tsukuba, 1-1-1 Tennodai, Tsukuba, 305-8571, Japan}
\newcommand{\KYOTO}{Institute for Chemical Research, Kyoto University, Gokasho, Uji-city, Kyoto 611-0011, Japan}

\newcommand{\beq}{\begin{equation}}
\newcommand{\eeq}{\end{equation}}
\newcommand{\beqa}{\begin{eqnarray}}
\newcommand{\eeqa}{\end{eqnarray}}

\begin{document}
\title{Hybrid quantum magnetic field sensor with an electron spin
and a nuclear spin in diamond}
\author{Yuichiro Matsuzaki}
   \affiliation{
\NTT
   }
 \author{ Takaaki Shimooka
 }
 \affiliation{\KYOTO}
 \author{Hirotaka Tanaka}
   \affiliation{
    \NTT
   }
    \author{Yasuhiro Tokura}
   \affiliation{
    \NTT
   }   \affiliation{
    \TSUKUBA
   }
       \author{Kouichi Semba}
   \affiliation{
    \NICT
   }
    \author{Norikazu Mizuochi
 }
 \affiliation{\KYOTO}

\begin{abstract}
Recently, magnetic
field sensors based on an electron spin of a nitrogen vacancy (NV)
 center in diamond have been studied
 both from an experimental and theoretical point of
 view. This system provides a nanoscale magnetometer, and it is possible to detect
 a precession of a single spin. In this paper, we propose a sensor
 consisting of an electron spin and a nuclear spin in diamond.
 Although the electron spin has a reasonable interaction  strength with magnetic
 field, the coherence time of the spin is relatively short.
 On
 the other hand, 
 the nuclear spin has a longer life time while the spin has a negligible
 interaction with magnetic fields. We show that, through the
 combination of such two different spins via the hyperfine interaction, it is possible to construct a
 magnetic field sensor with the sensitivity far beyond that of  previous
 sensors using just a single electron spin.
\end{abstract}

\maketitle


Measurement of the weak magnetic field with high spatial resolution
is an important objective in the field of metrology.
Many sensitive magnetic field sensor such as
SQUIDS \cite{simon1999local}, hall sensors in semiconductors
\cite{chang1992scanning}, and force sensors \cite{poggio2010force} have
been developed.
Also, a magnetic field sensor using entanglement has been also studied both from an experimental and theoretical point of
 view \cite{matsuzaki2011magnetic,chin2012quantum,huelga1997improvement,jones2009magnetic}.
One of the goals in this field is to measure a nuclear spin, because
of a wide variety of potential applications in many fields such as
material science and biomedical science.

Especially, much effort is being devoted to use
nitrogen-vacancy (NV) centers
for the realization of the field sensor to detect a single spin
\cite{maze2008nanoscale, taylor2008high, balasubramanian2008nanoscale, schaffry2011proposed}.
NV defects in diamond consist of a nitrogen atom
 and a vacancy in the adjacent site, which substitute for  carbon.
 Single qubit gates and readout of the spins in NV centers have already
 been demonstrated \cite{Go01a,gruber1997scanning,jelezko2002single,JGPGW01a}.
 There is an optical transition
 between its electron spin triplet ground state and 
 a first excited spin triplet state \cite{Go01a} in an NV center, and the quantum state of the electron spin can be
 measured via the fluorescence emission which has a dependency on the
 electron spin state
 \cite{gruber1997scanning,jelezko2002single}. 
 Also, Rabi oscillations of single electron spins in NV centers have been
 observed by using the optical detection \cite{JGPGW01a}.
 All these properties are prerequisite
in the construction of a sensitive and high-resolution sensor.

In this paper, we propose a scheme to improve the sensitivity of the  NV center
sensor by using a hybrid system of an electron spin and nuclear spin.
Here, each spin has its own distinct advantages. An electron spin offers
strong interactions with magnetic field, and therefore can efficiently mediate the
information of the magnetic field to the other system. A nuclear spin
presents excellent isolation from the environment, and this spin works as a
quantum memory to store the information.
Interestingly, NV center provide both an electron spin and a nuclear spin (${}^{13}$C,
${}^{14}$N, or ${}^{15}$N), and these spins are coupled via a hyperfine coupling.
We show that, by using this hybrid system, it is possible to detect
static magnetic fields
 with the sensitivity far beyond that of  previous
 sensors using just a single electron spin.

An NV center in
diamond has a spin 1 with the three levels $|0\rangle _{\text{e}}$ and $|\pm
1\rangle _{\text{e}}$, and it is possible to use just two of them to construct a two-level system,
namely a qubit.  The NV center has a zero-field splitting to be along the axis between the nitrogen
and the vacancy.
We apply an external magnetic field parallel to this axis, and
we set Zeeman splitting between 
states $|1\rangle _{\text{e}}$ and $|-1\rangle _{\text{e}}$.
This detuning allows us to use
an effective two level system. Throughout this paper, we assume to use
$|0\rangle _{\text{e}}$ and $|1\rangle _{\text{e}}$ as a two-level system to construct a
magnetic field sensor.

Let us summarize the conventional strategy to use just an electron spin
of the NV center for the detection of the magnetic field \cite{huelga1997improvement}.
In the present description, we make the assumption of no decoherence for simplicity.
Firstly, we prepare a superposition of the spin
$|+\rangle _{\text{e}}=\frac{1}{\sqrt{2}}|0\rangle _{\text{e}}
+\frac{1}{\sqrt{2}}|1\rangle _{\text{e}}$.
Secondly, let this spin expose a magnetic field for a time
$t$ , and we obtain $ |\psi (t)\rangle
_{\text{e}}=\frac{1}{\sqrt{2}}|0\rangle _{\text{e}}
+\frac{1}{\sqrt{2}}e^{-i\omega t}|1\rangle _{\text{e}}$
where $\omega $ denotes the detuning due to the Zeeman splitting induced
by the target magnetic field.
Finally, we measure the state $|\psi \rangle _{\text{e}}$ in the basis of
$\hat{\sigma }_y=|1_y\rangle _{\text{e}}\langle 1_y|-|0_y\rangle _{\text{e}}\langle 0_y|$
where $|1_{y}\rangle _{\text{e}}
=\frac{1}{\sqrt{2}}|0\rangle _{\text{e}}+\frac{i}{\sqrt{2}}|1\rangle
_{\text{e}}$ and $|0_{y}\rangle _{\text{e}}
=\frac{1}{\sqrt{2}}|0\rangle _{\text{e}}-\frac{i}{\sqrt{2}}|1\rangle
_{\text{e}}$. Note that we can construct a projection about $\hat{\sigma }_y $ by
rotating the spin with microwaves before the optical fluorescence measurement.
By repeating the above three processes $M$ times, we can obtain a
probability to project the state $|\psi (t)\rangle _{\text{e}}$ into
$|1_{y}\rangle _{\text{e}}$
as $P=\frac{1}{2}-\frac{1}{2}\sin \omega t$.
The uncertainty of the estimated value is then given by
\begin{eqnarray}
 |\delta \omega |=\frac{1}{|\frac{dP}{d\omega
  }|}\sqrt{\frac{P(1-P)}{M}}=\frac{1}{\sqrt{M}t}.
\end{eqnarray}
Therefore, for a longer exposure time of the sensor to the field, the
uncertainty of the estimated value becomes smaller.

However, in the actual circumstance, the noise from the environment
induces decoherence, and the non-diagonal term of the quantum state
disappears in a finite time. Typically, a relaxation time of the
electron spin in the NV center is much longer than the dephasing time
\cite{BMGP85a}, and so we consider only dephasing through this paper.
Since it is necessary to measure the state within the lifetime of the
 electron spin, we set the exposure time as $t= \alpha T^*_{2\text{e}}$
 where $T_{2\text{e}}^*$ denotes a
dephasing time of the electron spin and $\alpha $ denotes a small constant.
So the uncertainty is approximately calculated as
$ |\delta \omega |\simeq \frac{1}{\alpha \sqrt{M}T^*_{2\text{e}}}$. This
 shows that the uncertainty of the field sensing is limited by the
 lifetime of the electron spin.

 Unfortunately, there is a trade-off
 relation ship between the sensitivity and the spatial resolution of the
 field sensor. In order to achieve a spatial resolution, one needs to use
 a smaller nanocrystal. However, the miniaturization of the
crystal typically leads to the degradation of the coherence time of the
 electron spin in the diamond \cite{maze2008nanoscale,maletinsky2012robust,tisler2009fluorescence}. 

 We introduce our scheme to overcome the short lifetime of the
 electron spin by using a nuclear spin in the diamond.
A nuclear spin
is well isolated from the environment, and so we can use this as a
quantum memory to store the information. Actually, the coherence time of
 the nuclear spin in the NV center exceeds 1 second at room temperature
 by using the spin echo
 \cite{maurer2012room}.
 Instead, the coupling of the nuclear spin with the target magnetic
 field is three order of magnitude smaller than that of the electron
 spin. Fortunately, since the nuclear spin is coupled with an electron spin
 via a hyperfine coupling, it is possible to transfer the information attained
 by the electron spin to the nuclear spin for the storage. Actually, a
 controlled-not (C-NOT) gate between the electron spin and the nuclear
 spin has been already demonstrated \cite{NMRHWYJGJW01a}.
 Thus, we can
 construct an efficient hybrid magnetic field sensor to combine the
 preferable properties of these two
 different systems.

 The Hamiltonian of the NV center with an electron spin and a nuclear
 spin is described as follows.
The Hamiltonian is
\begin{eqnarray}
 &&H=D\hat{S}_{z,\text{e}}^2 +g^{(\text{e})}\mu
  _B(B_{\text{ex}}+B)\hat{S}_{z,\text{e}}
    +A\hat{S}_{z,\text{e}}\hat{\sigma }_{z,\text{n}}
  \nonumber \\
  &+&\frac{A'}{2}(\hat{S}_{+,\text{e}}\hat{\sigma }_{-,\text{n}}+\hat{S}_{-,\text{e}}\hat{\sigma
  }_{+,\text{n}})
  +g^{(n)}\mu
  _B(B_{\text{ex}}+B)\hat{\sigma }_{z,\text{n}},\ \ \ \ 
\end{eqnarray}
where $\omega =g\mu _BB$, $g^{(e)}$ ($g^{(n)}$), $\mu _B$, and $B$ ($B_{\text{ex}}$) denotes the
 Zeeman splitting, an electron (nuclear) spin
 g-factor, Bohr magneton, and target (external) magnetic field, respectively. 
 Since the Zeeman splitting of the nuclear spin due to the target
 magnetic  magnetic field is much weaker than the other values, we
 ignore this effect. Also, flip-flops between the electron spin and the
 nuclear spin can be neglected because of the energy difference between
 them. In addition, the Zeeman splitting induced by the external
 magnetic field allows us to detune $|-1\rangle _{\text{e}}$ and to isolate a
 two-level subsystem spanned by $|0\rangle _{\text{e}}$ and $|1\rangle
 _{\text{e}}$.
 So we obtain the following effective Hamiltonian.
 \begin{eqnarray}
 &&H\simeq (D +g^{(\text{e})}\mu
  _B(B_{\text{ex}}+B))|1\rangle _{\text{e}}\langle 1|
    +A|1\rangle _{\text{e}}\langle 1|\otimes \hat{\sigma
    }_{z,\text{n}}\nonumber \\
  &+&g^{(n)}\mu
  _BB_{\text{ex}}\hat{\sigma }_{z,\text{n}}.\ \ \ \ 
\end{eqnarray}
We make a unitary transformation
\begin{eqnarray}
 U=e^{i(D +g^{(\text{e})}\mu
  _BB_{\text{ex}})|1\rangle _{\text{e}}\langle 1| +g^{(n)}\mu
  _BB_{\text{ex}}\hat{\sigma }_{z,\text{n}}t},
\end{eqnarray}
 into a rotating frame,
  and this yield the following Hamiltonian in the frame
  \begin{eqnarray}
 &&H'\simeq g^{(\text{e})}\mu
  _BB|1\rangle _{\text{e}}\langle 1|
    +A|1\rangle _{\text{e}}\langle 1|\otimes \hat{\sigma
    }_{z,\text{n}}.
\end{eqnarray}

             \begin{figure}[h]
       \begin{center}
        \includegraphics[width=8cm]{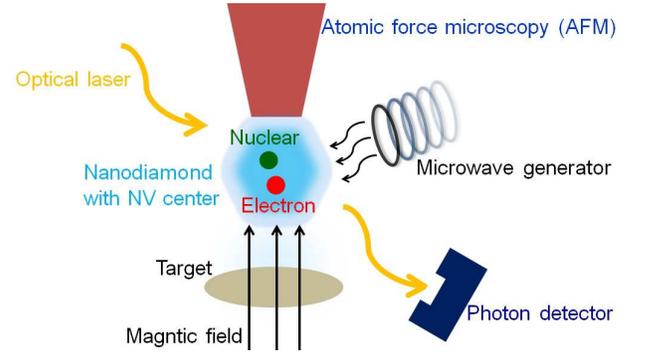} 
        \caption{
    The structure of our hybrid NV center sensor: a diamond containing an NV center with an
        electron spin and a nuclear spin is attached to an
        AFM-tip. Single qubit rotations and a C-NOT gate can be
        performed by directing microwave into the diamond. The
        electron-spin state can be measured by the optical laser and
        photodetectors.
        The electron spin has a reasonable interaction with the target
        magnetic field, and the nuclear spin works as a quantum memory
        to store the information from
        the magnetic field. By combining these two system, it is
        possible to improve the sensitivity of the field sensor.
        }
       \end{center}
       \end{figure}

\begin{figure}[h]
       \begin{center}
        \includegraphics[width=8.7cm]{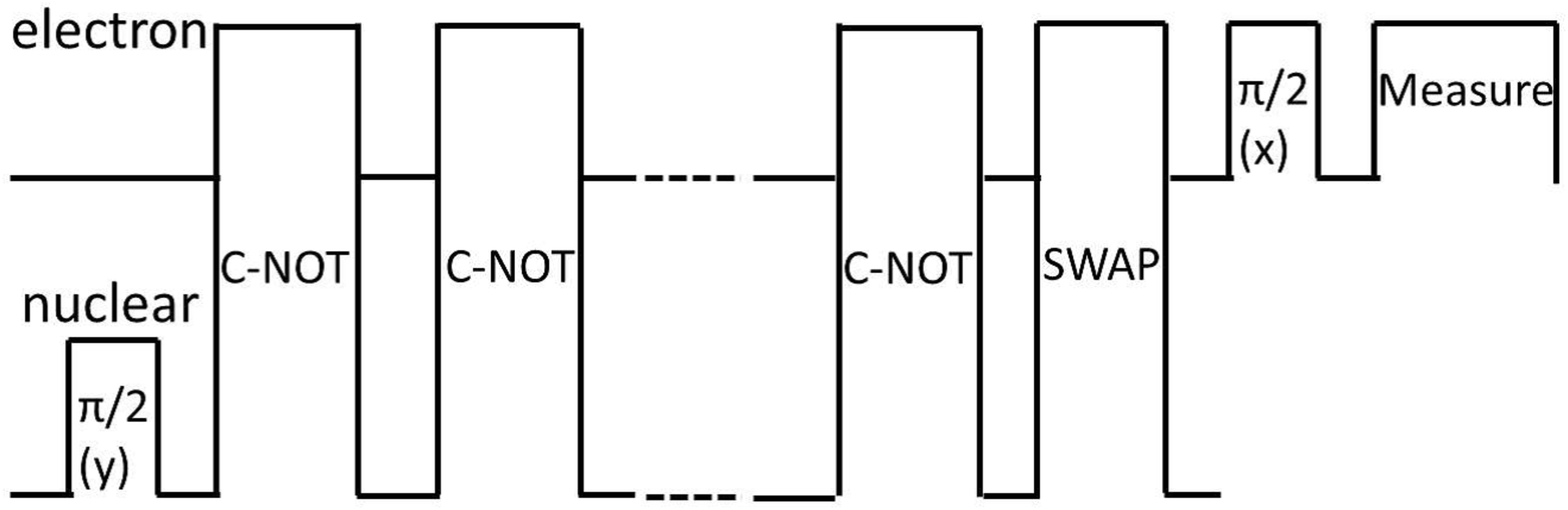} 
        \caption{A pulse sequence to detect magnetic fields with an
        electron spin and a nuclear spin in the NV center. In this
        sequence, we perform a
        single qubit rotation around $y$ axis, C-NOT gates, a SWAP gate,
        a single qubit rotation around $x$ axis, and optical detections. 
        We acquire the phase information from the coupling between the
        electron spin and magnetic field, and transfer the information
        to the nuclear-spin state. We perform this transfer $N$ times
        by implementing C-NOT gates $2N$ times,
        and finally readout the accumulated phase information.
        }\label{pulse}
       \end{center}
       \end{figure}

       We describe the prescription to detect the target magnetic field by our field
       sensor (see Fig. \ref{pulse}). For simplicity, we assume no decoherence for both
       of the electron spin and the nuclear spin.
       Firstly, we prepare $|0\rangle _\text{e}\otimes (\frac{1}{\sqrt{2}}|0\rangle _{\text{n}}
+\frac{1}{\sqrt{2}}|1\rangle _{\text{n}})$. Secondly, we perform a C-NOT
       gate between them where the
       electron is the target and the nuclear spin is the control, and
       we obtain $\frac{1}{\sqrt{2}}|0\rangle _{\text{e}}|0\rangle _{\text{n}}
+\frac{1}{\sqrt{2}}|1\rangle _{\text{e}}|1\rangle _{\text{n}}$. Thirdly,
       let this state evolves under the effect of the target magnetic
       field for a time $t=k T^*_{2\text{e}}$ and
       we obtain $\frac{1}{\sqrt{2}}|0\rangle _{\text{e}}|0\rangle _{\text{n}}
+\frac{1}{\sqrt{2}}e^{-i\omega t}|1\rangle _{\text{e}}|1\rangle
       _{\text{n}}$ where $k$ and $T^*_{2\text{e}}$ denotes
       a constant number and the dephasing time of the electron
       spin, respectively. Also, we define $\omega =g^{(\text{e})}\mu
  _BB +A $ as a resonant frequency of the state $|11\rangle
       _{\text{en}}$.
       Fourthly, we perform the C-NOT gate again to obtain a separable
       state $|0\rangle _\text{e}\otimes (\frac{1}{\sqrt{2}}|0\rangle _{\text{n}}
+\frac{1}{\sqrt{2}}e^{-i\omega t}|1\rangle _{\text{n}})$ where we
       transfer the phase
       information acquired by the electron spin to the nuclear spin.
By repeating the above four processes $N$ times, we obtain
       $|0\rangle _\text{e}\otimes (\frac{1}{\sqrt{2}}|0\rangle _{\text{n}}
+\frac{1}{\sqrt{2}}e^{-iN\omega t}|1\rangle _{\text{n}})=|0\rangle _\text{e}\otimes (\frac{1}{\sqrt{2}}|0\rangle _{\text{n}}
+\frac{1}{\sqrt{2}}e^{-ikN\omega T^{*}_{2\text{e}}}|1\rangle _{\text{n}})$. We measure
       the nuclear spin in $\hat{\sigma }_y$ basis which can be
       constructed by a SWAP gate between the electron spin and nuclear spin, a rotation of the electron spin, and
       the optical detection. Thus, we obtain the information
       of the phase $\theta =kN\omega T^{*}_{2\text{e}}$ stored in the superposition. Since the lifetime of
       the nuclear spin is much longer than that of the electron spin,
       it is possible to transfer the phase information several times
       from the electron spin to the nuclear spin before the
       non-diagonal term of the nuclear
       spins disappears. By repeating 
        such transfer, we can increase the amount of the phase accumulated
       from the target magnetic field, which enhances the
       sensitivity of the field sensor. 
       
       However, if we consider the effect of decoherence, there is of course a difficulty with this simple
picture, namely a propagation of the error from the electron spin to the
nuclear spin.  Due to the dephasing effect of the electron spin, the non-diagonal
term of the entangled state 
       $\frac{1}{\sqrt{2}}|0\rangle _{\text{e}}|0\rangle _{\text{n}}
+\frac{1}{\sqrt{2}}|1\rangle _{\text{e}}|1\rangle _{\text{n}}$
decreases as quickly as that of the electron spin does.
This dephasing error might be accumulated in the nuclear
spins, which could destroy the phase information obtained from the
target magnetic field.

Especially, if the dephasing noise is Markovian, the sensitivity of the
hybrid field sensor is as small as that of the
conventional one, due to the error propagation. In the Markovian
dephasing model, the non-diagonal term of the density matrix decays
exponentially, and so the error
probability to have a phase flip during the free evolution is calculated as
$\epsilon =\frac{1-e^{-\frac{t}{T^*_{2\text{e}}}}}{2}\simeq
\frac{k}{2}$ for $k\ll 1$.
When we implement the phase-information transfer from the electron spin
to the nuclear spin $N$ times, the total probability to have a phase
flip on the nuclear spin is calculated as $N\epsilon =\frac{Nk}{2}$.
In order to suppress the dephasing effect, we need a condition as $Nk<
1$. However, the acquired phase information from the target magnetic
field in this case is $\theta =kN\omega T^{*}_{2\text{e}}< \omega
T^{*}_{2\text{e}}$ which is comparable as that of the
conventional field sensor. So we cannot obtain any improvement of the
hybrid sensor in this case.

Fortunately, since the relevant dephasing in the NV center is
induced by low-frequency noise \cite{mizuochi2009coherence,
de2010universal} which is not Markovian, we
can suppress the error accumulation as follows. 
Under the effect of
low frequency noise, the non-diagonal term of the density matrix decays
quadratically. Due to this property, the initial decay of the non-Markovian noise
is slower than that of the Markovian noise. The error
probability to have a phase flip during the free evolution
is calculated as
$\epsilon =\frac{1-e^{-(\frac{t}{T^*_{2\text{e}}})^2}}{2}\simeq
\frac{k^2}{2}$ for $k\ll 1$. We need a condition $N\frac{k^2}{2}< 1$ to
suppress the dephasing effect after the $N$ times transfer, and so the scaling
of $k$ should be $k\propto \frac{1}{\sqrt{N}}$.
Thus, we have the acquired phase information from the target magnetic
field as $\theta =kN\omega T^{*}_{2\text{e}}\propto \sqrt{N}\omega
T^{*}_{2\text{e}}$, which can be larger as we increase the number of
the transfer. 
Therefore, we can improve the sensitivity of the hybrid magnetic field.

We perform more rigorous calculation to show the efficiency of our
magnetic field sensor. The relevant noise in this scheme is the dephasing on
the electron spin during the free evolution with the target magnetic
field, and so we only consider this error.
The density matrix after the $N$ times transfer can be described as follows
\begin{eqnarray}
 \rho &=&\frac{1}{2}|00\rangle _{\text{en}}\langle 00|+\frac{e^{Ni\omega t-N(\frac{t}{T^{*}_{2e}})^2}}{2}|00\rangle
  _{\text{en}}\langle 11|\nonumber \\
  &+&
  \frac{e^{-iN\omega t-N(\frac{t}{T^{*}_{2e}})^2}}{2}|11\rangle
  _{\text{en}}\langle 00|+\frac{1}{2}|11\rangle _{\text{en}}\langle 11|,
\end{eqnarray}
Here, since the coherence time of the nuclear spin is much longer than
that of the electron spin, we ignore the decoherence on the nuclear spin
\cite{maurer2012room}.
We set
$t=\frac{\alpha }{\sqrt{N}}T^*_{2e}$
where $\alpha $ denotes a constant number.
We can calculate the uncertainty of the estimated value as
\begin{eqnarray}
 |\delta \omega |=\frac{e^{\alpha ^2}}{\alpha} \frac{  1}{\sqrt{M}}\frac{\sqrt{1-e^{-2\alpha ^2}\sin ^2(\alpha
  \sqrt{N}\omega T^*_2)}}{
   \sqrt{N}T^*_2|\cos (\alpha \sqrt{N}\omega T^*_2)|
  },
\end{eqnarray}
where $M$ denotes the number of
the repetition of the experiment.
Since we try to detect a weak magnetic field, it is valid to assume
$\alpha \sqrt{N}\omega T^*_{2\text{e}}\ll 1$, and so we obtain
$ |\delta \omega |\simeq \frac{e^{\alpha ^2}}{\alpha} \frac{  1}{\sqrt{M}}\frac{1}{
   \sqrt{N}T^*_2
  }$.
Thus, the minimum uncertainty is attained for $\alpha
=\frac{1}{\sqrt{2}}$ as
$ |\delta \omega |_{opt}=\sqrt{2}e^{\frac{1}{2}}\cdot \frac{1}{\sqrt{M}}\frac{1}{\sqrt{N}T^*_{2\text{e}}}.$
On the other hand, the minimum uncertainty in the conventional scheme is
calculated as 
$|\delta \omega _{\rm{conv}}|_{opt}=\sqrt{2}e^{\frac{1}{2}}\frac{1}{\sqrt{M}T^*_{2e}}$
for $t=\frac{1}{\sqrt{2}}T^*_{2e}$ and $\omega T^*_{2e}\ll 1$
\cite{huelga1997improvement}.
Therefore, the uncertainty of our hybrid sensor is $\sqrt{N}$ times
smaller than that of the conventional one.

Note that our scheme can be interpreted as an application of quantum
Zeno effect \cite{BECG01a} to quantum metrology. Quantum Zeno effect
(QZE) is one of fascinating phenomena where a decay process is
suppressed by performing frequent projective measurements. It is known
that QZE can be observed if the survival probability $P_{\text{s}}(t)$
shows a quadratic decay as $P_{\text{s}}(t)\simeq 1-\Gamma ^2t^2$ for
$\Gamma t\ll 1$ where $\Gamma $ denotes a decay rate of the system. When
one performs $N$ projective measurements with a time interval $\tau
=\frac{t}{N}$, the success probability of projecting the state
in the excited level for all $N$ measurements is $P(t,N)\simeq (1-\Gamma
^2\tau ^2)^N\simeq 1-\Gamma ^2 \frac{t^2}{N}$, and therefore one can
increase the success probability as one increases the number of the
measurements.
It is known that QZE effect occurs in a system to show a quadratic decay while
QZE cannot be observed for exponential decay process
\cite{BECG01a}. Interestingly, we use the similar concept to construct a
hybrid magnetic field sensor. In our case, the free-evolution time of
the sensor is set to be in a time region when the decay process is quadratic,
and the effect of the dephasing of the electron spin is suppressed.

However, to observe a quadratic decay
behavior after each transfer process, we need to eliminate a correlation
between the system and the environment. In QZE, projective measurements
play a role of this resetting process, and so one can observe a quadratic decay
of the system
after each measurement. In contrast, we cannot eliminate the
correlation by measuring our system, because such projective
measurement destroys the phase information acquired by the magnetic
field.
This means that we need to wait until the correlation between the electron spin and
the environment disappears.
The correlation time of the environment around the NV spins has been
measured as $\tau _c\simeq 25 \ \mu s$  via dynamical decoupling
experiment \cite{de2010universal}. Therefore, our scheme involves a
waiting time $\tau _{\text{w}}$ which should be sufficiently larger than $25\ \mu s$ .

We discuss use cases of our approach using hybrid system.
Our approach could be relatively slow compared with the conventional approach.
The necessary time for a single cycle of the detection is limited by the
long waiting time as $t_{\text{our}}\simeq N\tau _{\text{w}}$.
 In
 contrast, in the conventional scheme, the time for a single cycle is $t_c=(\tau
_{\text{p}}+\frac{1}{\sqrt{2}}T^*_{2\text{e}}+
\tau 
_{\text{M}})$ where $\tau _{\text{p}}$, and
 $\tau _{\text{M}}$ denotes a preparation time of the initial state,
 and a measurement time, respectively. The typical time scales of $\tau
 _{\text{p}}$, $T^*_{2\text{e}}$, and
 $\tau _{\text{M}}$ are a few micro seconds, hundreds of nano seconds, and a few micro
 seconds \cite{NMRHWYJGJW01a,tisler2009fluorescence,de2010universal}.
 Since one needs to wait for $\tau _{\text{w}}$ at each transfer process,
 it takes longer for a single cycle in our scheme than the conventional
 one.
 If we fix the total time $T$ for the sensing and try to minimize the uncertainty of the
 estimated value, our sensor may not be so sensitive as the conventional
 one, because $M$ becomes smaller. 
 However, if we fix the number of measuring the spin $M$ and try to
 minimize the uncertainty, our sensor is
 superior to the conventional one. Actually, this is the case when we
 use this magnetic field sensor at low temperature or on a
 photo-sensitive materials such as  biological tissues \cite{ono2013entanglement}. In order to readout the
 electron spin, it is necessary to irradiate the optical laser which
 generates heat and could damage the surface of the materials to be
 measured. In such circumstance, we need to restrict the number of
 measurements $M$ to avoid heating or damage \cite{ono2013entanglement}.
  In our sensor, we can decrease the uncertainty of the estimated value
   by transferring the information from the electron spin to the nuclear
   spin.

    \begin{figure}[h!] 
\begin{center}
\includegraphics[width=0.90 \linewidth]{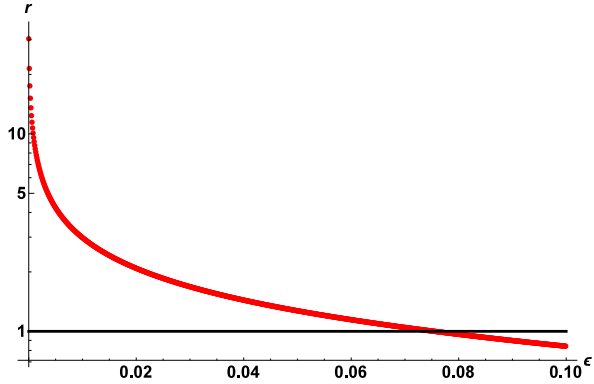} 
\caption{
 We plot a relative sensitivity $r=\frac{|\delta \omega
   _{\rm{conv}}|_{opt}}{|\delta \omega |_{opt}}$ against a gate error
 $\epsilon $ where $|\delta \omega |_{opt}$ ($|\delta \omega
   _{\rm{conv}}|_{opt}$) denotes the sensitivity of our scheme (the
 conventional scheme). 
 If the gate error is below $0.1\%$, the
 sensitivity of our sensor is one order of magnitude better than the
 conventional one. 
 }
 \label{optimize}
\end{center}
\end{figure}
Imperfection of gate operations are the primary source of errors in our
scheme.
Especially, we perform $2N$ C-NOT gates and one SWAP gate, whose imperfection would
decrease the efficiency of our scheme.
Suppose that we are subject to depolarizing noise. Here, the
state becomes an identity operator with a probability $\epsilon $ when
we implement the two-qubit gate. Since we perform $(2N+1)$ two-qubit
gates, we obtain the following state after the final SWAP gate
\begin{eqnarray}
  \rho _N
 &=&(1-\epsilon )^{2N+1} (\frac{1}{2}|0\rangle _{\text{e}}\langle
  0|+
  \frac{e^{i\alpha \sqrt{N}\omega T^*_{2e}-\alpha ^2}}{2}|0\rangle
  _{\text{e}}\langle 1|\nonumber \\
  &+&\frac{e^{-i\alpha \sqrt{N}\omega T^{*}_{2e}-\alpha ^2}}{2}|1\rangle
  _{\text{e}}\langle 0|+\frac{1}{2}|1\rangle _{e}\langle 1|)\otimes |0\rangle _{\text{n}}\langle 0|
  \nonumber \\
  &+&\{1-(1-\epsilon
  )^{2N+1}\} \frac{\hat{\openone}_{\text{en}}}{4}.
\end{eqnarray}
If we have $\alpha \sqrt{N}\omega T^*_2\ll 1$, we obtain
$\delta  \omega  \simeq \frac{1}{(1-\epsilon )^{2N+1}e^{-\alpha ^2}\alpha
   \sqrt{N}T^*_2}\frac{1}{\sqrt{M}}$. This means that,
 if the error rate of each two-qubit gate operation is below $0.1\%$, we
   can perform hundreds of such gate operations without significant
   degradation of the fidelity so that
  the uncertainty can be order of magnitude smaller than that of the
  conventional one. Since the sensitivity for the conventional scheme is
   calculated as $|\delta \omega
   _{\rm{conv}}|_{opt}=\sqrt{2}e^{\frac{1}{2}}\frac{1}{\sqrt{M}T^*_{2e}}$,
   the ratio between them is calculated as $r=\frac{|\delta \omega
   _{\rm{conv}}|_{opt}}{|\delta \omega |}=\sqrt{2}e^{\frac{1}{2}}(1-\epsilon )^{2N+1}e^{-\alpha ^2}\alpha
   \sqrt{N} $. For a given $\epsilon $, we can maximize $r=\frac{|\delta \omega
   _{\rm{conv}}|_{opt}}{|\delta \omega |_{opt}}$ to choose
   the optimum $N$ and $\alpha $, and plot this in the
   Fig. \ref{optimize}.
   As long as the error rate is below $7.5\%$, we can achieve an
   enhancement over the conventional strategy.

 In conclusion, we propose a hybrid magnetic field sensor using an
 electron spin and a nuclear spin. The electron spin strongly interacts with the
 target magnetic field while the nuclear spin has a long coherence
 time. We have found that, by combining the best of two worlds, we can
 constructs an efficient magnetic field sensor with a sensitivity far
 beyond that of a simple NV center.

 During preparation of this manuscript, we became aware
of a related work of quantum sensing with a quantum memory \cite{zaiser2016enhancing}.

 This work was supported by JSPS KAKENHI Grants
15K17732
 and MEXT KAKENHI Grant Number 15H05870.

\end{document}